\documentclass[%
 reprint,
 amsmath,amssymb,
 aps,
 longbibliography,
pre,
prl
]{revtex4-2}

\usepackage[caption=false]{subfig}
\usepackage{graphicx}
\usepackage{dcolumn}
\usepackage{bm}
\usepackage{xcolor}
\usepackage{gensymb}
\usepackage{soul}
\usepackage{comment}
\usepackage{siunitx}
\usepackage {hyperref}
\hypersetup{
    colorlinks = true,
    linkcolor = blue,
    anchorcolor = blue,
    citecolor = blue,
    filecolor = blue,
    urlcolor = blue
    }
\DeclareMathAlphabet{\mathpzc}{OT1}{pzc}{m}{it}

\begin{document}

\preprint{APS/123-QED}

\title{Communication: Modeling layered mosaic perovskite alloy microstructures across length scales via a packing algorithm}

\author{Murray Skolnick}
\affiliation{
    Department of Chemistry, Princeton University, Princeton, New Jersey 08544, USA
}%

\author{Salvatore Torquato}
\email{torquato@electron.princeton.edu}
\affiliation{
    Department of Chemistry, Department of Physics, Princeton Institute for the Science and Technology of Materials, and Program in Applied and Computational Mathematics, Princeton University, Princeton New Jersey 08544, USA
}%

\date{\today}

\begin{abstract}

Layered ``mosaic" metal-halide perovskite materials display a wide-variety of microstructures that span the order-disorder spectrum and can be tuned via the composition of their constituent \emph{B}-site octahedral species. 
Such materials are typically modeled using computationally expensive \textit{ab initio} methods, but these approaches are greatly limited to small sample sizes. 
Here, we develop a highly efficient hard-particle packing algorithm to model large samples of these layered complex alloys that enables an accurate determination of the geometrical and topological properties of the \emph{B}-site arrangements within the plane of the inorganic layers across length scales. 
Our results are in good agreement with various experiments, and therefore our algorithm bypasses the need for full-blown \textit{ab initio} calculations. 
The accurate predictive power of our algorithm demonstrates how our minimalist hard-particle model effectively captures complex interactions and dynamics like incoherent thermal motion, out of plane octahedral tilting, and bond compression/stretching. 
We specifically show that the composition-dependent miscibility predicted by our algorithm for certain silver-iron and copper-indium layered alloys are consistent with previous experimental observations. 
We further quantify the degree of mixing in the simulated structures across length scales using our recently developed \textcolor{black}{sensitive} ``mixing" metric. 
The large structural snapshots provided by our algorithm also shed light on previous experimentally measured magnetic properties of a copper-indium system. 
The generalization of our algorithm to model 3D perovskite alloys is also discussed. 
In summary, our packing model and mixing metric enable one to accurately explore the enormous space of hypothetical layered mosaic alloy compositions and identify materials with potentially desirable optoelectronic and magnetic properties. 

\end{abstract}

\maketitle

\begin{figure}
    \centering
    \subfloat[\label{fig:single}]{\includegraphics[width=0.5\columnwidth]{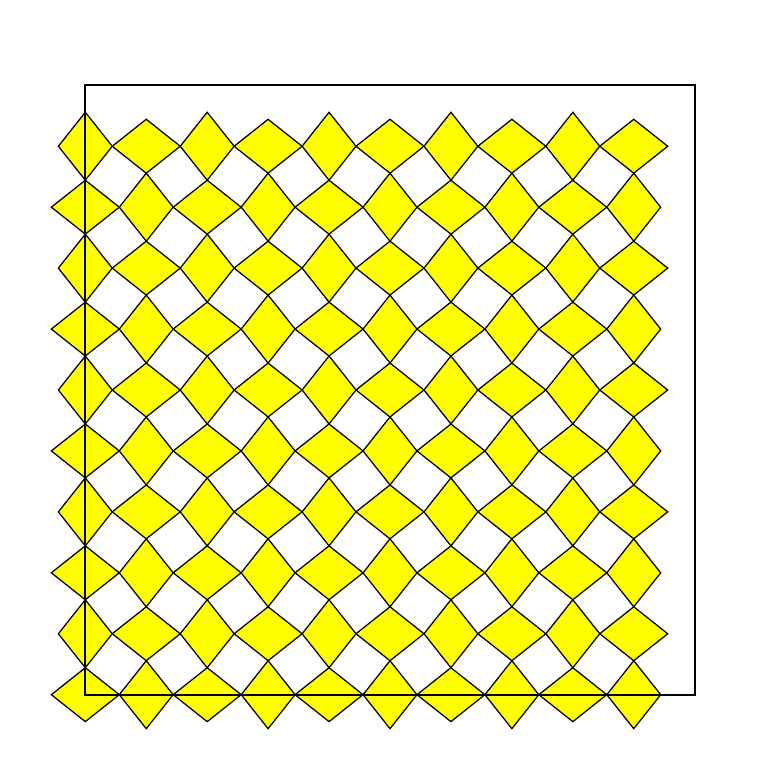}}%
    \subfloat[\label{fig:double}]{\includegraphics[width=0.5\columnwidth]{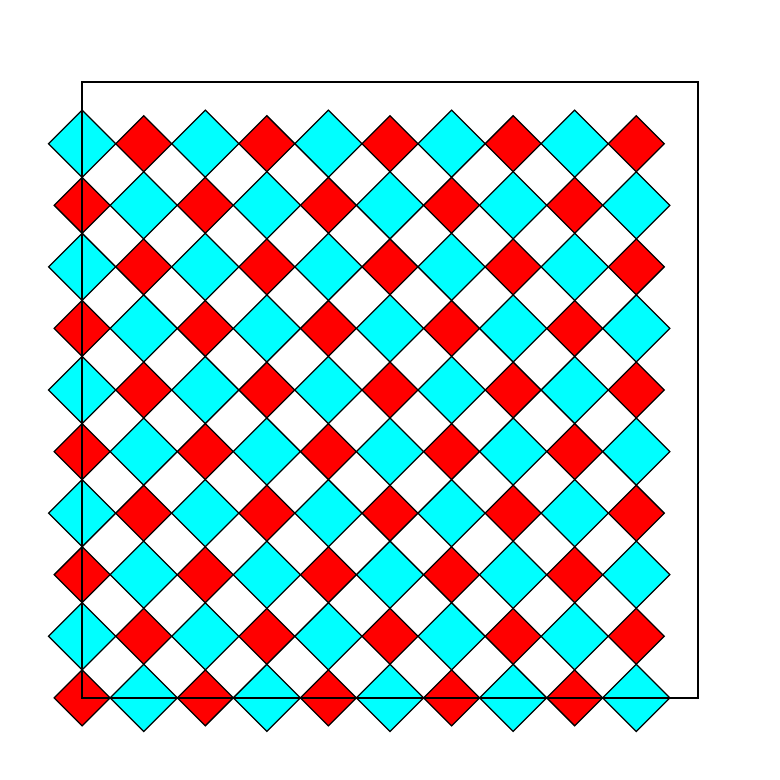}}
    \caption{Images of rhombi arrangements generated by the LASC algorithm for an individual layer of the (a) single perovskite $\mathrm{A_2Cu^{II}Cl_4}$ and (b) double perovskite $\mathrm{A_4Cu^{I}In^{III}Cl_8}$. In both (a) and (b), the cross sections of the Jahn-Teller distorted \cite{Co21} $\mathrm{Cu^{II}Cl_6}$, $\mathrm{Cu^{I}Cl_6}$, and $\mathrm{In^{III}Cl_6}$ units within the plane of the layer are respectively depicted by yellow, cyan, and red rhombi, and the solid black square denotes the boundary of the periodic simulation cell. The centers of the rhombi are located on the pervoskite \emph{B}-sites. The \emph{A}-sites are centered on the interstitial spaces between the \emph{B}-sites but are located out of the plane of the layer. Note that the vertices of the aforementioned rhombi correspond to the positions of the $\mathrm{Cl}$ ligands. The antiferrodistortive and rock-salt arangements of (a) and (b), respectively, are consistent with those observed in the experimentally obtained crystal structures for these materials reported in Ref. \cite{Co21}. For the simulation of (a), we fix the Jahn-Teller elongated axis of $\mathrm{Cu^{II}Cl_6}$ in the plane of the layer as is observed experimentally \cite{Co21}. The structures in (a) and (b) consist of $N=100$ rhombi total with $L=10$ rhombi per side.}\label{fig:crystals}
\end{figure}

Layered ``mosaic" halide perovskite alloys are an emerging class of materials with varying degrees of order/disorder across length scales. 
They specifically incorporate three different $\mathrm{MX}_6$ (M=transition metal, X=halide) octahedra at the \emph{B}-site within their inorganic layers, which are separated by spacer molecules ($\textrm{A}$) that occupy the perovskite \emph{A}-site \cite{De25, Co21, Li23, Vi25}. 
In the schematic cross sections of the inorganic layers shown in Fig. \ref{fig:crystals}, the centers of the octahedra (or rhombi within the plane) are located on the \emph{B}-sites. The \emph{A}-sites are located on the interstitial spaces between the \emph{B}-sites and are out of the plane. 
Layered mosaic alloys are prepared by combining precursor layered \textit{crystalline} single and double perovskites, whose structures respectively contain one and two different $\mathrm{MX}_6$ species, either in solution \cite{Co21} or via mechanochemical ball-milling \cite{Li23,Vi25}. 
Schematic images of the idealized arrangement of $\mathrm{Cu^{II}Cl_6}$ octahedra in a layer of the single perovskite $\mathrm{A_2Cu^{II}Cl_4}$, and $\mathrm{Cu^{I}Cl_6}$ and $\mathrm{In^{III}Cl_6}$ octahedra in a layer of the double perovskite $\mathrm{A_4Cu^{I}In^{III}Cl_8}$ are shown in Figs. \ref{fig:single} and \ref{fig:double}, respectively. 

Note that the electrons of transition metal atoms of the three different $\mathrm{MX_6}$ octahedra at the \emph{B}-site of the mosaic alloys can interact with one another within the inorganic layers via intervalence charge transfer \cite{Co21,Li23} and magnetic superexchange coupling \cite{Ka59, Be15a, Da79}. 
Therefore, the optoelectronic and magnetic properties of mosaic alloys are directly related to and can be tuned by the degree of mixing of the different octahedra at the \emph{B}-site across different length scales \cite{Co21,Li23,Vi25}. Specifically, a high degree of mixing at length scales corresponding to the lattice spacing can result in a high degree of intervalence charge transfer \cite{Li23}. By contrast, greater phase-segregation of ferromagnetically coupled species across short and intermediate length scales can lead to interesting bulk magnetic properties \cite{Vi25}. 
Therefore, the capacity to efficiently and accurately predict the \emph{B}-site arrangement of a specific combination of starting materials and quantify its degree of mixing enhances our fundamental understanding of the formation and physical properties of mosaic alloys. 

Conventional atomistic modeling of the formation of these alloys, which entails breaking and forming chemical bonds, would require computationally expensive \textit{ab initio} molecular dynamics methods \cite{Ma09b,Zh19b} in order to adequately capture important interactions and dynamics like thermal motion, out of plane octahedral tilting, and bond stretching/compression \cite{Ma09b,Zh19b}.  However, the large computational time and resource costs of such calculations place limitations on the size and number of systems that one can practically simulate. To circumvent such computational limitations, we posit that the aforementioned physics governing the formation of these materials can be adequately captured by a pseudo hard-particle packing algorithm. This simplification is motivated by the observation by Connor \textit{et al.} that the mixing behaviors of the layered mosaic alloy $\mathrm{A_4(Cu^{I}In^{III}) Cu^{II}_2 Cl_8}$ is likely a direct consequence of the need to minimize packing mismatches between its constituent differently sized and shaped corner-sharing octahedra within the inorganic layers. 
Such packing constraints are highlighted in the crystalline structures \ref{fig:single} and \ref{fig:double}, in which the octahedra (or rhombi in the plane) are arranged such that there are no gaps or overlaps (mismatch) between their shared nearest-neighbor vertices. Note that while the idealized crystals \ref{fig:single} and \ref{fig:double} have precisely no mismatch between neighboring vertices, it is expected that sufficiently small vertex-vertex mismatch corresponds to natural octahedral tilting and distortion that occurs in 2D perovskites \cite{Co20,Ma19b,Bi19b}. Thus, we will incorporate such effects in our packing model as detailed below. 

We specifically adapt here the computationally efficient Adaptive Shrinking Cell (ASC) stochastic packing algorithm \cite{To09c,To09b,At12,Ma21}---an established protocol for packing hard-particles in Euclidean space---to the case of hard rhombi fixed to a square lattice in order to efficiently model the local tiling patterns of the \emph{B}-site octahedra across various length scales within an individual inorganic layer. 
This algorithm, which we refer to as the Lattice Adaptive Shrinking Cell (LASC) algorithm, finds optimal arrangements of the square lattice-bound \emph{B}-site octahedra subject to the constraint of minimizing packing mismatch between the shared vertices of nearest-neighbor octahedra within an inorganic layer. 
We show via our recent mixing metric \cite{Sk24} \textcolor{black}{(see Eq. \eqref{eqn:sigma} for definition)}, previously introduced to characterize mixing in heterogeneous media, that the LASC algorithm predicts phase separation and mixing behaviors in certain copper-indium and silver-iron alloys that are in qualitative agreement with experimental observations. 
Additionally, we show via percolation analysis that the bulk magnetic properties of the LASC predictions for the mosaic alloy $\mathrm{A_4 (Cu^{I}In^{III})Cu^{II}_2 Cl_8}$ are consistent with experimental measurements. 
Overall, our successful application of the LASC algorithm to these copper-indium and silver-iron alloys further highlight its efficiency and accurate predictive power, as well as its utility in traversing the sizable parameter space of potential alloy chemical compositions and stoichiometries\textcolor{black}{--especially when used in conjunction with our mixing metric \eqref{eqn:sigma}.} 

\begin{figure*}[t]
    \centering
    \subfloat[\label{fig:interactions}]{\includegraphics[width=0.8\textwidth]{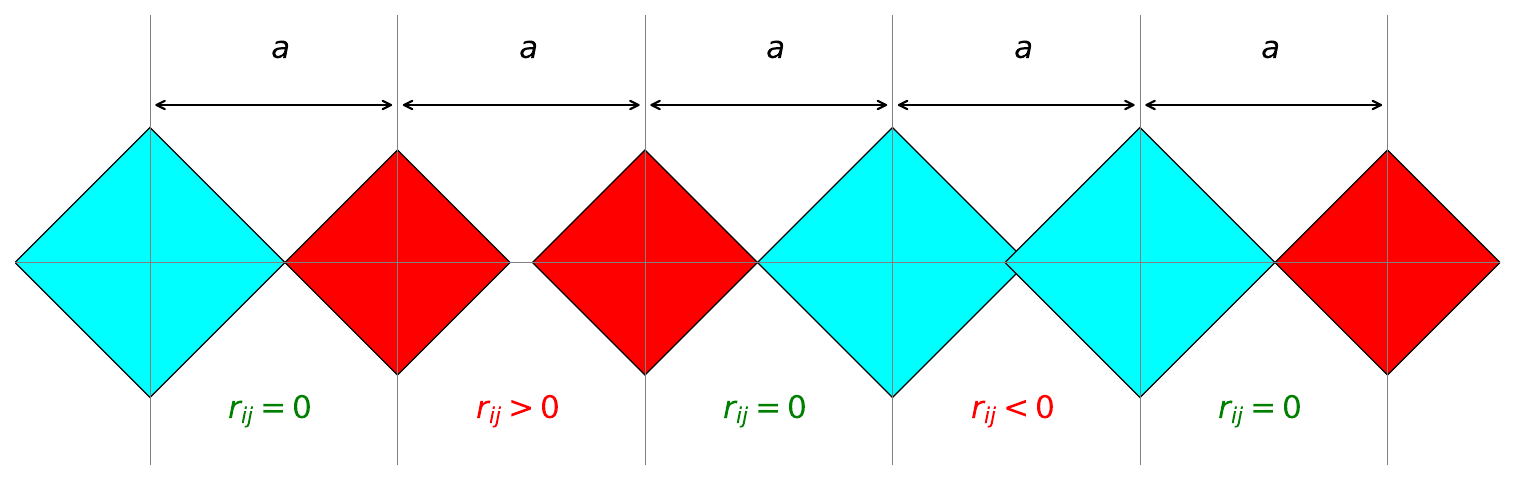}}\\
    \subfloat[\label{fig:alg}]{\includegraphics[width=0.8\textwidth]{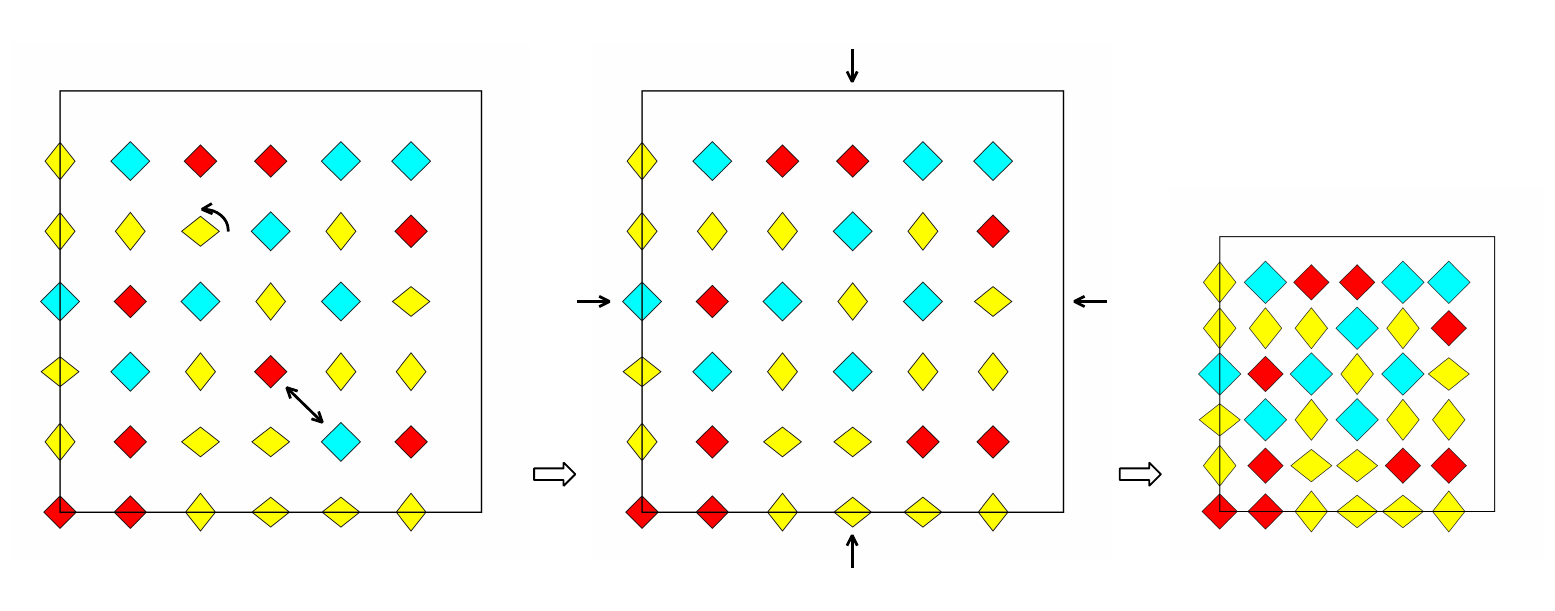}}
    \caption{(a) A schematic depicting a row within a layer of the double perovskite crystal $\mathrm{A_4Cu^{I}In^{III}Cl_8}$ during a LASC simulation in which the nearest-neighbor rhombi interact via potential \eqref{eqn:pot}. Here, the \textcolor{black}{third and fourth rhombi from the left} have been swapped, resulting in an unfavorable gap $r_{ij}>0$ between the nearest-neighbor vertices of the adjacent $\mathrm{In^{III}Cl_6}$ units \textcolor{black}{(second and third from the left)}, and an unfavorable overlap $r_{ij}<0$ between the nearest-neighbor vertices of the adjacent $\mathrm{Cu^{I}Cl_6}$ units \textcolor{black}{(fourth and fifth from the left)} within the plane of the layer. By contrast, the nearest-neighbor $\mathrm{Cu^{I}Cl_6}$ and $\mathrm{In^{III}Cl_6}$ units have no such packing mismatches (i.e., $r_{ij}=0$) because their $\mathrm{M-Cl}$ bond lengths are consistent with the lattice constant $a$. (b) A schematic depiction of the (left) trial Monte Carlo swap and rotational moves, (middle) isotropic cell compression moves employed by the LASC algorithm, as well as the (right) resulting densified system. In (a), the gray lines represent the grid of the underlying square lattice. In (b), the magnitude of the trial compression is much greater than what is normally applied in the algorithm to facilitate visualization.}
    \label{fig:schematics}
\end{figure*}

We now describe the details of the LASC algorithm. 
The \emph{B}-site octahedra are treated as non-deformable rhombi with \textcolor{black}{in plane} diagonal lengths $l_p$ ($p=1,2$) that are fixed to an $L\times L$ site square lattice (lattice constant $a$). 
Henceforth, we denote the distance (in {\AA}) between the adjacent vertices of the rhombi at lattice sites $i$ and $j$ as 
\begin{equation}
    r_{ij}=a - \frac{l_p^{(i)}+l_p^{(j)}}{2}\label{eqn:rij_def},
\end{equation}
\textcolor{black}{where} $l_p^{(i)}$ and $l_p^{(j)}$ are the respective diagonal lengths of the rhombi at nearest-neighbor sites $i$ and $j$ \textcolor{black}{that are parallel to the line connecting sites $i$ and $j$.}
Note that $r_{ij}$ is positive (negative) for gaps (overlaps). 
We posit that small gaps and overlaps \textcolor{black}{($|r_{ij}|\lesssim0.02l_p^{(i/j)}$)} between nearest-neighbor vertices correspond to normal incoherent thermal motion and vibrational modes \cite{Co21}, while larger \textcolor{black}{($|r_{ij}|\gtrsim 0.05l_p^{(i/j)}$)} mismatches represent nonphysical bonds between neighboring octahedra or appreciable out of plane octahedral tilting \cite{Vi25}. 
Henceforth, we refer to these small and large gaps/overlaps as ``favorable" and ``unfavorable", respectively. 
To mimic these interactions in our minimalist LASC model, rhombi within the layer interact via nearest-neighbor vertex-vertex interactions through the effective pair-potential
\begin{equation}
    v(r_{ij}) = k\left( \left|\frac{r_{ij}}{\sigma}\right|^{\alpha} + \left|\frac{r_{ij}}{\sigma}\right|^{\beta}\right )\label{eqn:pot},
\end{equation}
where the parameters $k>0$ and $\sigma>0$ respectively control the strength and characteristic length scale of the interaction, and the exponents $\alpha$ and $\beta$ are described below. 

The bond lengths (rhombus diagonals) $l_p$ used in the potential \eqref{eqn:pot} are obtained from crystallography experiments \cite{Co21,Li23b}. The specific values of $l_p$ for the \emph{B}-site units considered here are listed in Table \ref{tab:bondlengths}. 
The strict non-overlapping constraint of the original ASC scheme is relaxed in potential \eqref{eqn:pot} by allowing overlaps between neighboring rhombi at the cost of an energetic penalty. 
Unlike in true hard-particle systems, however, \textit{gaps} between neighboring vertices are also penalized by Eq. \eqref{eqn:pot}. 
Interactions between rhombi via potential \eqref{eqn:pot} are depicted schematically in Fig. \ref{fig:interactions}. 
As found through trial and error on the known benchmark crystal structures $\mathrm{A_2Cu^{II}Cl_4}$ and $\mathrm{A_4Cu^{I}In^{III}Cl_8}$ \cite{Co21}, severe mismatches (i.e., $|r_{ij}|\gtrsim l_p^{(i/j)}$) are penalized by taking $\beta\geq2$, while making $0<\alpha<1$ effectively penalizes mismatches with $r_{ij}\gtrsim0.02 l_p^{(i/j)}$\footnote{For simplicity, gaps and overlaps of equal magnitude $|r_{ij}|$ incur the same energetic penalty.}. 
Here, we take $\alpha=1/2$ and $\beta=2$ but other combinations (e.g., $\alpha=1/4$ and $\beta=6$) are expected to yield similar structures. 
We also observed empirically that taking $k=1$ arb. unit and $\sigma=1$ {\AA} in the potential \eqref{eqn:pot} for all possible rhombus-rhombus interactions resulted in correct reproduction of the aforementioned benchmark crystal structures \footnote{\textcolor{black}{Taking $\sigma=1$ {\AA} is physically reasonable, since the interactions captured by potential \eqref{eqn:pot} occur at distances that are of the order of the length of a chemical bond. By contrast, the parameter $k$ only sets the energy scale of the system, so we take it to be unity for simplicity.}}. 

In the LASC algorithm, the task of finding dense arrangements of the octahedra with minimal packing mismatches amounts to finding minima of the fictitious energy
\begin{equation}
    E = \sum_{\langle i,j \rangle} v(r_{ij})\label{eqn:etot}
\end{equation}
in which $\langle i,j \rangle$ denotes summation over all nearest-neighbor pairs within the periodic simulation super cell. 
The LASC scheme solves this energy minimization problem using simulated annealing via the following steps which are depicted schematically in Fig. \ref{fig:alg}. An initial random (uncorrelated) mixture at a low density ($a\sim10l_p$) at the desired stoichiometry is densified through a series of $N_c$ isotropic and irreversible cell compression steps. 
Subsequently, each lattice site is subject to random trial swap moves a total of $N_s$ times which attempt to remove packing mismatches introduced by the compression steps. Since octahedra with Jahn-Teller distortions within the plane can adopt two different orientations, these rhombi are subject to additional random discrete $90\degree$ rotational moves as another way to remove packing mismatches.

The aforementioned trial moves are all accepted according to the Boltzmann criterion, i.e.,
\begin{equation}
    p_{{\rm acc}}({\rm old}\to {\rm new}) = \min\left\{1, \exp\left( -\frac{E_{{\rm new}} - E_{{\rm old}}}{T} \right) \right\}\label{eqn:pacc}
\end{equation}
where $E_{{\rm old}}$ and $E_{{\rm new}}$ are the energy \eqref{eqn:etot} before and after the move and $T$ is the fictitious annealing temperature. 
The initial temperature $T_0$ is set high (i.e., $T_0>>k$) such that $p_{{\rm acc}}\sim0.5$ early on in order to adequately explore the configurational space and prevent the system from getting stuck in a shallow local energy minimum. As the simulation progresses, the annealing temperature is gradually lowered according to the power law cooling schedule $T(n)=\gamma^n T_0$ in which $n$ is the annealing stage and $\gamma\in(0,1)$ is the cooling factor ($\gamma=0.98$ is used here). This cooling schedule has the net effect of gradually lowering the acceptance probability \eqref{eqn:pacc} as the system converges to a low energy \textcolor{black}{(i.e., low packing mismatch)} state. Here, we take the number of compression and random swap and rotation moves to be one ($N_c,N_s=1$) for each simulation loop, and repeat until the scaled energy per lattice site $E/(L^2k)$ converges to a constant value \textcolor{black}{within the tolerance $\epsilon=10^{-3}$. Physically, this energy convergence behavior corresponds to the system approaching an energetic / packing mismatch minimum.}
As specific benchmark tests for LASC, we reproduce the known crystal structures \cite{Co21} of the inorganic layers of the single perovskite $\mathrm{Cu^{II}Cl_4}$ and double perovskite $\mathrm{Cu^{I}In^{III}Cl_8}$ which are shown in Figs. \ref{fig:single} and \ref{fig:double}, respectively. 

\begin{table}
\caption{\label{tab:bondlengths} Experimentally obtained values of $\textrm{X-M-X}$ bond lengths within the plane of an inorganic layer which we use to parameterize the diagonals $l_p$ of the octahedra (rhombi) in the LASC simulations.}
\begin{ruledtabular}
\begin{tabular}{cccc}
$\mathrm{MX_6}$ & $l_1$ ({\AA}) & $l_2$ ({\AA}) \\ 
\colrule
$\mathrm{Cu^{I}Cl_6}$ \cite{Co21} & 6.00 & 6.00 \\ 
$\mathrm{In^{III}Cl_6}$ \cite{Co21} & 5.00 & 5.00 \\ 
$\mathrm{Cu^{II}Cl_6}$ \cite{Co21} & 5.80 & 4.60 \\ 
$\mathrm{Ag^{I}Cl_6}$ \cite{Li23b} & 5.64 & 5.64 \\ 
$\mathrm{Fe^{III}Cl_6}$ \cite{Li23b} & 4.74 & 4.74 \\ 
$\mathrm{Fe^{II}Cl_6}$ \cite{Li23b} & 5.10 & 5.10 \\ 
\end{tabular}
\end{ruledtabular}
\end{table}

We also probe the mixing characteristics of the individual $\mathrm{MX_6}$ octahedra in the LASC structures using our recently developed metrics for quantifying the degree of phase mixing and separation in multi-phase heterogeneous materials across length scales \cite{Sk24}. 
Specifically, we compute the local \textit{molar-fraction variance}, $\sigma_{\chi_i}^2(R)$, of species $i$ associated with a circular sampling window of radius $R$. Note that $\sigma_{\chi_i}^2(R)$ is a decreasing function of $R$ for which the upper bound is $\sigma_{\chi_i}^2(R=0)=\chi_i(1-\chi_i)$ and $\lim_{R\to\infty}\sigma_{\chi_i}(R)=0$. As an overall quantification of mixing across length scales in the entire system, we compute here the integrated compositional mixing metric \cite{To22a,Ma24a}
\begin{equation}
    \Sigma_{\chi_i} = \frac{1}{\chi_i(1-\chi_i)} \int_0^{\infty} \sigma_{\chi_i}^2(R) dR \label{eqn:sigma},
\end{equation}
for each octahedra species $i$. The factor of $[\chi_i(1-\chi_i)]^{-1}$ in Eq. \eqref{eqn:sigma} normalizes the metric so that mixing in the alloys can be compared across compositions. Smaller (larger) values of $\sigma_{\chi_i}^2(R)$ and $\Sigma_{\chi_i}$ correspond to a greater degree of mixing (phase separation) across length scale $R$ and the entire system, respectively \cite{Sk24}.

\begin{figure}[h!]
    \centering
    \subfloat[\label{fig:AgFe_img}]{\includegraphics[width=0.8\columnwidth]{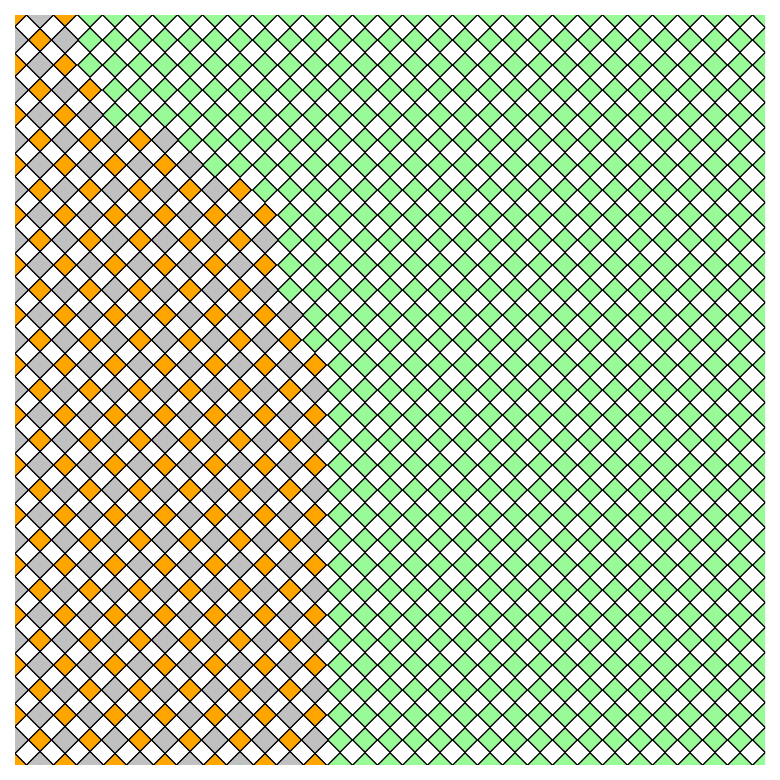}} \\
    \subfloat[\label{fig:AgFe_mix}]{\includegraphics[width=0.8\columnwidth]{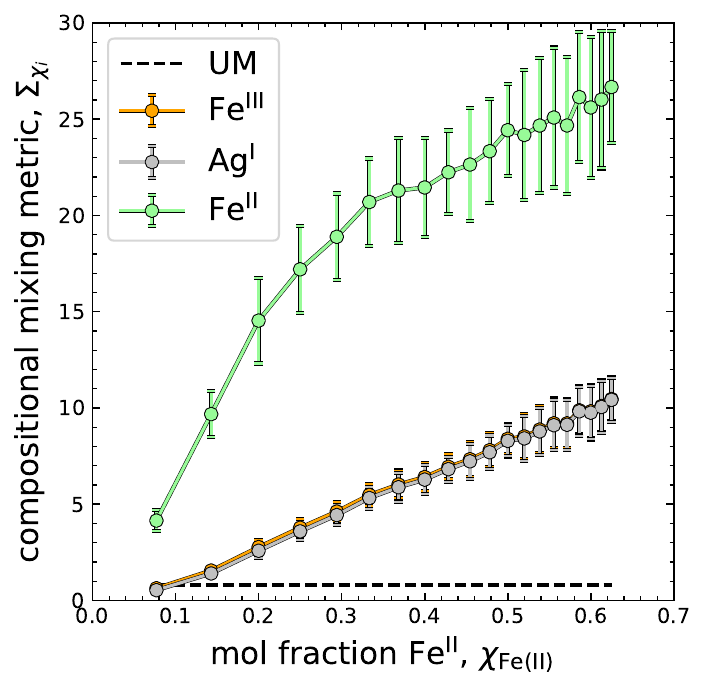}}
    \caption{(a) A portion of a LASC simulated $\mathrm{(Ag^{I}Fe^{III}) Fe^{II}_2}$ structure consisting of phase segregated domains of $\mathrm{Fe^{II}Cl_4 }$ and $\mathrm{Ag^{I}Fe^{III}Cl_8}$. (b) Plots of the ensemble averaged integrated compositional mixing metric \eqref{eqn:sigma} as a function of $\chi_{\mathrm{Fe(II)}}$ for $\mathrm{Fe^{III}Cl_6}, \mathrm{Ag^{I}Cl_6},$ and $\mathrm{Fe^{II}Cl_6}$ octahedra. The dashed black line marks the value of the mixing metric for an uncorrelated \emph{B}-site mixture (UM), which one can show is $\approx0.8033$ for all compositions $0<\chi_i<1$ \cite{To22a}. Note that $\Sigma_{\mathrm{Fe(III)}}$ and $\Sigma_{\mathrm{Ag(I)}}$ are roughly equivalent across $\chi_{\mathrm{Fe(II)}}$ values since these octahedra are always co-localized in the rock-salt pattern of the double perovskite $\mathrm{Ag^{I}Fe^{III}Cl_8}$. In (a), the $\mathrm{Fe^{III}Cl_6}$, $\mathrm{Ag^{I}Cl_6}$, and $\mathrm{Fe^{II}Cl_6}$ octahedra within the layer are respectively represented as orange, gray, and green rhombi. In (b), the error-bars represent the standard deviation of the ensemble. Each composition is averaged over 250 realizations of a $200\times200$ lattice system.}
    \label{fig:AgFe_results}
\end{figure}

As the first non-trivial test of our algorithm, we generate candidate structures of the hypothetical layered mosaic alloys $\mathrm{A_{4(x/2+y)} (Ag^{I}Fe^{III})_y Fe^{II}_x Cl_{8(x/2+y)}}$ (henceforth referred to as $\mathrm{(Ag^{I}Fe^{III})_y Fe^{II}_x}$). 
Experimentally, it was found that ball-milling single perovskite $\mathrm{A_2Fe^{II}Cl_4}$ with double perovskite $\mathrm{A_4 Ag^{I}Fe^{III}Cl_8}$ did not produce the mosaic $\mathrm{(Ag^{I}Fe^{III})_y Fe^{II}_x}$---rather, the starting materials remained phase separated \cite{Li23b}. 
Indeed, LASC simulations of the alloy $\mathrm{(Ag^{I}Fe^{III})_y Fe^{II}_x}$ across $\mathrm{Fe^{II}}$ molar fractions $0.1\lesssim \chi_{\mathrm{Fe(II)}}\lesssim 0.6$ (i.e., $1\leq x/(x + 2y)\leq20$) produce structures in which the starting materials remain phase separated, as seen in Fig. \ref{fig:AgFe_img}. 
Such phase segregation behavior is supported quantitatively by the compositional mixing metric $\Sigma_i$ $(i=\mathrm{Fe^{III}},\mathrm{Ag^{I}},\mathrm{Fe^{II}})$, which is between $\sim3\times$ and $\sim25\times$ larger---or more phase separated---than the metric for a random (uncorrelated) \emph{B}-site mixture \footnote{In such a mixture, each \emph{B}-site within the layer is occupied by species $i$ with probability $\chi_i$. One can computationally generate such uncorrelated mixtures by randomly shuffling the arrangement of octahedra in a LASC structure.} at the same bulk composition (see Fig. \ref{fig:AgFe_mix} and the corresponding caption for further details). 

\begin{figure}[h!]
    \centering
    \subfloat[\label{fig:CuIn_img}]{\includegraphics[width=0.8\columnwidth]{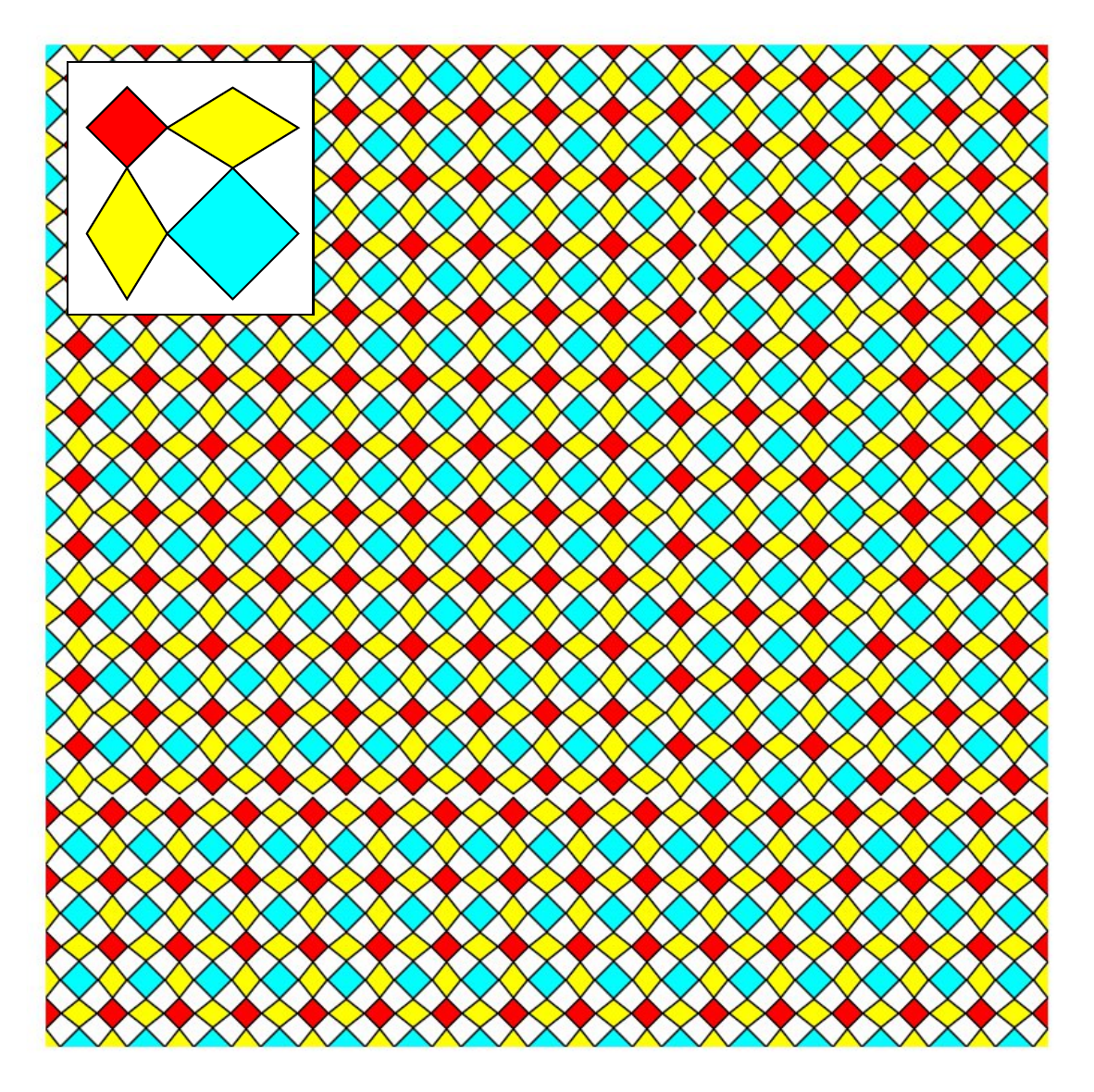}} \\
    \subfloat[\label{fig:CuIn_mix}]{\includegraphics[width=0.8\columnwidth]{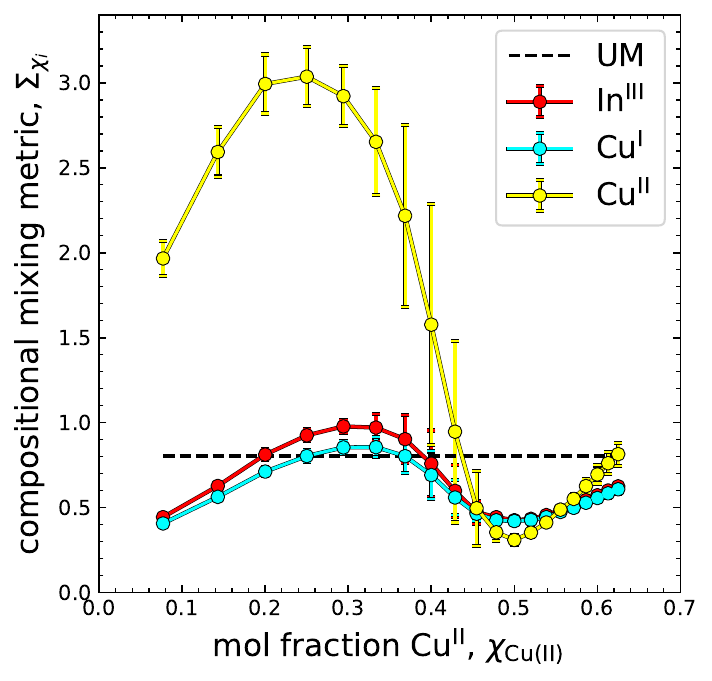}} 
    \caption{(a) An example image of a LASC simulated $\mathrm{(Cu^{I}In^{III}) Cu^{II}_2}$ structure, in which polycrystalline grains of the $2\times2$ motif shown in the inset are observed. Note the horizontal and vertical grain boundaries near the bottom and right of the structure, respectively. (b) Plots of the ensemble averaged integrated compositional mixing metric \eqref{eqn:sigma} as a function of $\chi_{\mathrm{Cu(II)}}$ for $\mathrm{In^{III}Cl_6}, \mathrm{Cu^{I}Cl_6},$ and $\mathrm{Cu^{II}Cl_6}$ octahedra. The dashed black line marks the value of the mixing metric for an uncorrelated \emph{B}-site mixture (UM). In (a), the $\mathrm{In^{III}Cl_6}$, $\mathrm{Cu^{I}Cl_6}$, and $\mathrm{Cu^{II}Cl_6}$ octahedra within the layer are respectively represented as red, cyan, and yellow rhombi. In (b), the error-bars represent the standard deviation of the ensemble. Each composition is averaged over 250 realizations of a $200\times200$ lattice system.}
    \label{fig:CuIn_results}
\end{figure}

Given these successful predictions by the LASC algorithm, we now tackle the more challenging structural prediction of the experimentally observed \cite{Co21,Li23} layered mosaic alloy $\mathrm{A_4 (Cu^{I}In^{III})Cu^{II}_2 Cl_8}$ (henceforth referred to as $\mathrm{(Cu^{I}In^{III}) Cu^{II}_2}$).
Previous experimental observations showed that combining the single perovskite $\mathrm{A_2 Cu^{II}Cl_4}$ and double perovskite $\mathrm{A_4Cu^{I}In^{III}Cl_8}$ in a variety of proportions $x:y$ always lead to the formation of the $1:2$ alloy $\mathrm{(Cu^{I}In^{III}) Cu^{II}_2}$, with diffraction experiments indicating that $\mathrm{(Cu^{I}In^{III}) Cu^{II}_2}$ exhibited disorder at the \emph{B}-site, and magnetic studies demonstrating that this alloy is paramagnetic, despite the starting material $\mathrm{A_2 Cu^{II}Cl_4}$ being ferromagnetic \footnote{Note that the ferromagnetism in $\mathrm{A_2 Cu^{II}Cl_4}$ is attributable to in-plane exchange interactions between nearest-neighbor $\mathrm{Cu^{II}}$ ions \cite{Co21}.}.  
These observations lead to the proposal in Ref. \cite{Co21} of a tiling motif (see inset of Fig. \ref{fig:CuIn_img}) that minimizes packing mismatch, and is consistent with the observed stoichiometry of and lack of long-range magnetic ordering \footnote{In this motif, the $\mathrm{Cu^{II}}$ ions are \textit{next}-nearest-neighbors. Therefore, they do not form a sample spanning (percolating) ferromagnetic cluster.} in the mosaic alloy $\mathrm{(Cu^{I}In^{III}) Cu^{II}_2}$. 

Importantly, this \textit{exact} tiling motif is observed throughout the LASC simulated $\mathrm{(Cu^{I}In^{III}) Cu^{II}_2}$ alloy shown in Fig. \ref{fig:CuIn_img}. 
Moreover, domains of this repeating tile with different orientations are observed throughout the structure. 
This polycrystallinity, which was suggested in Ref. \cite{Co21} and observed in all of the simulated $\mathrm{(Cu^{I}In^{III}) Cu^{II}_2}$ alloys, provides a possible explanation for the experimentally observed \emph{B}-site disorder in $\mathrm{(Cu^{I}In^{III}) Cu^{II}_2}$ \cite{Co21}. 
We also computed the percolation thresholds for the network of ferromagnetically coupled nearest-neighbor $\mathrm{Cu^{II}}$ ions in the LASC structures via the ``burning algorithm" \cite{St92} across compositions $\chi_{\mathrm{Cu(II)}}$. 
Our calculations indicate that sample spanning clusters of \textcolor{black}{ferromagnetically coupled} $\mathrm{Cu^{II}}$ ions, \textcolor{black}{and thus long-range magnetic order \cite{Co21,St92}}, would only start to appear in hypothetical alloys $\mathrm{(Cu^{I}In^{III}) Cu^{II}_{x}}$ where $x> 20$, i.e., the threshold is estimated to be $\chi^{(p)}_{\mathrm{Cu(II)}}\approx0.91$. \textcolor{black}{Overall, this simulated percolation behavior is consistent with the simple paramagnetism observed in the real material $\mathrm{A_4 (Cu^{I}In^{III})Cu^{II}_2 Cl_8}$ \cite{Co21}.}

Quantification of mixing in the simulated copper-indium alloys as a function of the molar fraction of $\chi_{\mathrm{Cu(II)}}$ via the metric \eqref{eqn:sigma} indicates that the experimentally observed $\mathrm{(Cu^{I}In^{III}) Cu^{II}_2}$ alloy, in which $\chi_{\mathrm{Cu(II)}}=0.5$, is the most well-mixed among the simulated compositions. 
Specifically, the metric $\Sigma_i$ for all three species (i.e., $i=\mathrm{In^{III}},\mathrm{Cu^{I}},\mathrm{Cu^{II}}$) are simultaneously minimized at $\chi_{\mathrm{Cu(II)}}=0.5$. 
Moreover, they are about twice as well mixed as they would be in an uncorrelated \emph{B}-site mixture. 
Such optimal mixing is consistent with intervalence charge transfer and the associated emergent optical properties of the alloy $\mathrm{(Cu^{I}In^{III}) Cu^{II}_2}$ detailed in Refs. \cite{Co21,Li23}. 
Interestingly, the degree of packing mismatch in the LASC structures as measured by the scaled energy per site $ E/(L^2k)$ has a local minimum precisely at $\chi_{\mathrm{Cu(II)}}=0.5$, which is consistent with the experimentally observed preference toward formation of the $\mathrm{(Cu^{I}In^{III}) Cu^{II}_2}$ alloy.

In summary, we have demonstrated the high predictive power of the LASC algorithm for a variety of real layered mosaic alloys, and our results suggest that there is no need for full-blown \textit{ab initio} calculations to simulate the formation of mosaic alloys. 
\textcolor{black}{Via percolation thresholds, we showed that the simple paramagnetism of the $\mathrm{(Cu^{I}In^{III}) Cu^{II}_2}$ alloy is in quantitative agreement with experimental observations \cite{Co21}. 
Moreover, the respective phase-separation and mixing properties of the silver-iron and copper-indium alloys, as observed empirically in Figs. \ref{fig:AgFe_img} and \ref{fig:CuIn_img}, agree qualitatively with experiments \cite{Co21,Li23b}. 
This agreement with experiments is further bolstered by quantitative characterization of such behaviors in the simulated alloys via the mixing metric $\Sigma_i$. 
We note that obtaining such a robust quantification of mixing is challenging to achieve with conventional entropy-based mixing metrics, which we previously found to depend on the degree of coarse graining of the underlying sampling grid, resulting in difficulty distinguishing mixing across large-length scales \cite{Sk24}.} 


\textcolor{black}{Overall, these} successful predictions highlight how the simple, heuristic potential \eqref{eqn:pot} effectively captures the salient atomic interactions and dynamics of these materials, such as highly compressed or stretched unphysical bonds between neighboring octahedra, 3D effects like out of plane octahedral tilting, as well as random thermal motion. 
\textcolor{black}{We further} note that the LASC algorithm is computationally efficient as evidenced by the fact that it takes about 3.5 hours to generate a $200\times200$ lattice site structure on a single CPU core. 
By contrast, gaining a snapshot of the \emph{B}-site arrangement of a mosaic alloy across such length scales using \textit{ab initio} molecular dynamics simulations would require much greater time and computing resources \cite{Zh19b,Ma09b}. 
\textcolor{black}{Thus, combining the speed and predictive accuracy of the LASC algorithm with the capacity of the mixing metric $\Sigma_{\chi_i}$ to sensitively rank order the degree of mixing in alloys enables high-throughput screening of candidate mosaic alloys within the enormous parameter space of chemical compositions and stoichiometries.} 




We hypothesize that our simple packing algorithm works well for the following reasons: the strong repulsive interactions that are relevant in condensed phases of matter are often well captured by excluded-volume or similar interactions; see e.g., Ref. \cite{To18b} and references therein. 
Additionally, the compression steps of our packing algorithm effectively capture the mechanochemical alloying process used to synthesize mosaic alloys in the laboratory \cite{Li23,Vi25}. 
These results suggest that the LASC algorithm can be fruitfully applied to aid in the characterization of other mosaic alloys. 
Indeed, we will soon report in a comprehensive manuscript the productive application of the LASC algorithm and our mixing metrics $\sigma^2_{\chi_i}(R)$ and $\Sigma_{\chi_i}$ in combination with experiment to study a novel silver-chromium mosaic alloy and its magnetic properties \cite{Vi25}. 

While the results of this paper and our forthcoming manuscript \cite{Vi25} are concerned with layered mosaic perovskites, we note that the LASC algorithm is readily generalized to the problem of predicting 3D alloy structures that are fully three dimensional. 
Specifically, one would treat the $\mathrm{MX_6}$ units as non-deformable octahedra fixed to a cubic lattice and include an additional pair of nearest-neighbors in energy \eqref{eqn:etot}. 
We anticipate that LASC simulations could be particularly helpful in probing the bulk structures of 3D perovskites, whose structures tend to be more complicated than their layered 2D counterparts \cite{Li23c}. 

\begin{acknowledgments}
This paper was sponsored by the National Science Foundation under Award No. CBET-2133179, as well as the Army Research Office and was accomplished under Cooperative Agreement No. W911NF-22-2-0103. M.S. also acknowledges the support of the Harold W. Dodds Honorific Fellowship. The authors thank the Princeton Institute for Computational Science and Engineering for the computational resources.
\end{acknowledgments}


 \newcommand{\noop}[1]{}
%


\end{document}